\pdfoutput=1
\RequirePackage{ifpdf}
\ifpdf 
\documentclass[pdftex]{sigma}
\else
\documentclass{sigma}
\fi

\numberwithin{equation}{section}

\newtheorem{Theorem}{Theorem}[section]
\newtheorem*{Theorem*}{Theorem}
\newtheorem{Corollary}[Theorem]{Corollary}
\newtheorem{Lemma}[Theorem]{Lemma}

\theoremstyle{definition}

\newcommand{\ZZ}{\mathbb{Z}}
\newcommand{\og}{\mathrm{OG}}
\newcommand{\rg}{\mathrm{RG}}
\newcommand{\tr}{\operatorname{tr}}

\begin{document}

\allowdisplaybreaks

\newcommand{\arXivNumber}{2510.26137}

\renewcommand{\PaperNumber}{056}

\FirstPageHeading

\ShortArticleName{GUE Correlators and Large Genus Asymptotics}

\ArticleName{GUE Correlators and Large Genus Asymptotics}

\Author{Jiayi ZHAO}

\AuthorNameForHeading{J.~Zhao}

\Address{School of Mathematical Sciences, University of Science and Technology of China,\\
230026, P.R.~China}
\Email{\mail{zhaojiayi616@mail.ustc.edu.cn}}

\ArticleDates{Received December 19, 2025, in final form May 29, 2026; Published online June 03, 2026}

\Abstract{In this paper, we use a formula obtained by Dubrovin and Yang (2017) to study certain asymptotic behaviors of GUE (Gaussian unitary ensemble) correlators. More precisely, we obtain large genus asymptotics of enumerations of ordinary graphs and ribbon graphs with 1 face.}

\Keywords{GUE correlator; ribbon graph; ordinary graph; large genus asymptotics; matrix-resolvent formula}

\Classification{05A16; 14N10; 15B52}

\section{Introduction}
Large genus asymptotics of psi-class intersection numbers on the moduli space of stable algebraic curves have been studied in the literature. Liu and Xu found and proved~\cite{LX} certain large genus asymptotics and rationality properties. Delecroix, Goujard, Zograf and Zorich (DGZZ) conjectured~\cite{DGZZ} the large genus uniform leading asymptotics of psi-class intersection numbers, which was improved and proved by Aggarwal in~\cite{Agg}. Guo and Yang~\cite{GYPsi} gave a new proof of the above-mentioned Liu--Xu's results and DGZZ's conjecture through estimates of the matrix-resolvent formula given in~\cite{BDY1} regarding the KdV hierarchy.

In this paper, using the matrix-resolvent formulae~\cite{DYGUE} for the GUE (Gaussian unitary ensemble) correlators (which are closely related to enumerations of ribbon graphs), we will study certain large genus asymptotic behaviors and certain rationality properties. Our method is similar to~\cite{GYPsi} (cf.~\cite{BDY1,DYP1}). In particular, we will obtain large genus asymptotics of enumerations of ordinary graphs and ribbon graphs with 1 face. We note that Pasquetti and Schiappa~\cite{FreeEnergy} analyzed large genus behavior of certain GUE free energy from the perspective of resurgence theory; see also~\cite{Resurgence} for a comprehensive review on resurgence. We also note that a deep result about uniform bounds on enumeration of triangulations in large genus was obtained in~\cite{Triangular}.

The GUE correlators of observables $\tr M^i$, $i=1,2, \dots $, are defined by
\[
\bigl\langle\tr M^{i_1}\cdots\tr M^{i_n}\bigr\rangle(N):=\frac{\int_{\mathcal{H}(N)}\tr M^{i_1}\cdots\tr M^{i_n} {\rm e}^{-\frac{1}{2}\tr M^2} {\rm d}M}{\int_{\mathcal{H}(N)}{\rm e}^{-\frac{1}{2}\tr M^2}{\rm d}M},
\]
where $\mathcal{H}(N)$ denotes the space of $N\times N$ Hermitian matrices. It is known~\cite{AvM,DYGUE,Gera} that \emph{the GUE partition function}
\[
	Z_N(\textbf{s};\epsilon)=\frac{(2\pi)^{-N}\epsilon^{-1/12}}{\text{Vol}(N)}\int_{\mathcal{H}(N)}{\rm e}^{-\operatorname{tr}V(M)}{\rm d}M,
\]
\emph{where $\mathcal{H}(N)$ is the space of $N\times N$ Hermitian matrices and}
$V(M)=\frac{1}{2}M^2-\sum_{j\ge1}s_jM^j$, $\textbf{s}=(s_1,s_2, \dots )$,
\emph{is a tau-function of the Toda lattice hierarchy, with the time variables defined by $t_j=s_{j+1}/\epsilon$, $j=0,1,2, \dots $.}

The \emph{connected} GUE correlators \smash{$\bigl\langle\operatorname{tr}M^{i_1}\cdots\operatorname{tr}M^{i_n}\bigr\rangle_c(N)$} can be defined through the relation
\[
	\bigl\langle\operatorname{tr}M^{i_1}\cdots\operatorname{tr}M^{i_n}\bigr\rangle(N)=\sum_{\mathcal{P}\text{ partition of }\{1,\dots,n\}}\prod_{I\in\mathcal{P}}\biggl\langle\prod_{j\in I}\tr M^{i_j}\biggr\rangle_c(N).
\]
For $n\ge1$, denote the generating series for the $n$-point correlators by
\[
	C_n(N;\lambda_1,\dots,\lambda_n):=\sum_{i_1,\dots,i_n=1}^\infty\frac{\bigl\langle\operatorname{tr}M^{i_1}\cdots\operatorname{tr}M^{i_n}\bigr\rangle_c(N)}{\lambda_1^{i_1+1}\cdots\lambda_n^{i_n+1}}.
\]
Using the matrix-resolvent method, Dubrovin and Yang~\cite{DYGUE} obtained an explicit formula for $C_n(N;\lambda_1,\dots,\lambda_n)$,
which reads as follows:
\begin{align}
	&C_1(N;\lambda)=N\sum_{j\ge1}\frac{(2j-1)!!}{\lambda^{2j+1}}({_2F}_1(-j,-N;2;2)-j\cdot{_2F}_1(1-j,1-N;3;2)),\label{main1} \\
	&C_n(N;\lambda_1,\dots,\lambda_n)=-\frac{1}{n}\sum_{\sigma\in
		S_n}\frac{\tr\bigl(R_N\bigl(\lambda_{\sigma(1)}\bigr)\cdots R_N\bigl(\lambda_{\sigma(n)}\bigr)\bigr)-\delta_{n,2}}{\prod_{i=1}^n\bigl(\lambda_{\sigma(i)}-\lambda_{\sigma(i+1)}\bigr)},\qquad n\ge2,\label{main2}
\end{align}
where $S_n$ denotes the symmetric group, $\sigma(n+1):=\sigma(1)$, ${_2F}_1(a,b;c;z)$ is the
Gauss hypergeometric function
\[
	{_2}F_1(a,b;c;z)=\sum_{j=0}^\infty\frac{(a)_j(b)_j}{(c)_j}\frac{z^j}{j!}=1+\frac{ab}{c}\frac{z}{1!}+\frac{a(a+1)b(b+1)}{c(c+1)}\frac{z^2}{2!}+\cdots,
\]
and, for $N\ge1$, $R_N(\lambda)$ denotes the following matrix-valued series
\[
	R_N(\lambda):=
	\begin{pmatrix}
		1&0\\0&0
	\end{pmatrix}+N\sum_{j=0}^\infty\frac{(2j-1)!!}{\lambda^{2j+2}}
	\begin{pmatrix}
		(2j+1)A_{N,j}&-\lambda B_{N+1,j}\\
		\frac{\lambda}{N}B_{N,j}&-(2j+1)A_{N,j}
	\end{pmatrix},
\]
where
\[
	A_{N,j}:={_2F}_1(-j,1-N;2;2),\qquad B_{N,j}:={_2F}_1(-j,1-N;1;2).
\]
See~\cite{HZ,MS,Toda,Zhou} for relevant explicit formulae (cf. also~\cite{BE,FY}).

Denote by $\og_g(i_1,\dots,i_n)$ the weighted number of connected ordinary graphs of genus~$g$ with~$n$ vertices of valencies $i_1,\dots,i_n$. We know from Euler's formula $1-g=V-E$ with $V=n$, \smash{$E=\frac{i_1+\cdots+i_n}{2}$} and $g$ the first Betti number of the graph (number of loops) that
\[
	i_1+\cdots+i_n=2g+2n-2.
\]
Then we have
\[
	\og_g(i_1,\dots,i_n)=\frac{1}{i_1!\cdots i_n!n!}\bigl\langle\operatorname{tr}M^{i_1}\cdots\operatorname{tr}M^{i_n}\bigr\rangle_c(1).
\]
Introduce a normalization of the enumeration of ordinary graphs by
\[
	\mathcal{G}^\og_{i_1,\dots,i_n}(g):=\frac{i_1!\cdots i_n!n!}{(2g+2n-3)!!}\og_g(i_1,\dots,i_n).
\]
Certain large genus asymptotics for the enumeration of ordinary graphs as $g\to\infty$, as well as a structure involving linear combinations of finitely many exponentials for fixed number of vertices, are established in~\cite{DYZ}. We will prove in Section~\ref{secog} the following theorems on the large genus asymptotics and rationality for ordinary graphs.

\begin{Theorem}\label{thmOG}
	For fixed $n\ge1$ and fixed integers $i_1,\dots,i_{n-1}\ge1$, we have
	\[
		\lim_{g\to+\infty}\mathcal{G}^\og_{i_1,\dots,i_{n-1},2g+2n-2-|i|}(g)=1,
	\]
	where $|i|=i_1+\cdots+i_{n-1}$.
\end{Theorem}
\begin{Theorem}\label{thm2OG}
	For fixed $n\ge1$ and fixed integers $i_1,\dots,i_{n-1}\ge1$, there exists a rational function \smash{$R^{\og}(g;i_1,\dots,i_{n-1})$} of $g$ that may depend on $n,i_1,\dots,i_{n-1}$, such that
	\[
		\mathcal{G}^{\og}_{i_1,\dots,i_{n-1},2g+2n-2-|i|}(g)=R^{\og}(g;i_1,\dots,i_{n-1}).
	\]
\end{Theorem}
Theorems~\ref{thmOG} and~\ref{thm2OG} yield the following corollary.
\begin{Corollary}
	For fixed $n\ge1$ and fixed integers $i_1,\dots,i_{n-1}\ge1$, we have the following convergent expansion:
	\begin{equation}
		\mathcal{G}^\og_{i_1,\dots,i_{n-1},2g+2n-2-|i|}(g)\sim\sum_{k=0}^{\infty}\frac{\mathcal{G}^\og_k}{g^k},\qquad g\to\infty,\label{ogexp}
	\end{equation}
	where \smash{$\mathcal{G}^{\og}_0=1$} and \smash{$\mathcal{G}^{\og}_k=\mathcal{G}^{\og}_k(n,i_1,\dots,i_{n-1})$}, $k\ge1$, are constants.
\end{Corollary}

Denote by $a_g(i_1,\dots,i_n)$ the weighted number of connected ribbon graphs of genus~$g$ with~$n$ vertices of valencies $i_1,\dots,i_n$ (see~\cite{DYGUE}). Then Euler's formula for closed surfaces tells that $V-E+F=2-2g$, where $V=n$, \smash{$E=\frac{i_1+\cdots+i_n}{2}$}, and $F$ is the number of faces. Now we focus on graphs with 1 face. We denote the number of ribbon graphs of this kind by $\rg_g(i_1,\dots,i_n)$. Then Euler's formula implies that
\[
	i_1+\cdots+i_n=4g+2n-2.
\]
In fact, $\rg_g(i_1,\dots,i_n)$ is the first-order coefficient of $N$ in the polynomial
\[
\frac{1}{n!}\bigl\langle\operatorname{tr}M^{i_1}\cdots\operatorname{tr}M^{i_n}\bigr\rangle_c(N)
\]
 (see~\cite{graph, HZ}, cf.~\cite{DYGUE}), i.e.,
\[
	\bigl\langle\operatorname{tr}M^{i_1}\cdots\operatorname{tr}M^{i_n}\bigr\rangle_c=n!\rg_g(i_1,\dots,i_n)N+O\bigl(N^2\bigr).
\]
We introduce the normalized numbers of ribbon graphs as follows:
\[
	\mathcal{G}^{\rg}_{i_1,\dots,i_n}(g):=\frac{n!}{2\cdot(4g+2n-5)!!}\rg_g(i_1,\dots,i_n).
\]
We will prove in Section~\ref{secrg} the following two theorems.
\begin{Theorem}\label{thmRG}
	For fixed $n\ge1$ and fixed integers $i_1,\dots,i_{n-1}\ge1$, we have
	\[
		\lim_{g\to+\infty}\mathcal{G}^{\rg}_{i_1,\dots,i_{n-1},4g+2n-2-|i|}(g)=1,
	\]
	where $|i|=i_1+\cdots+i_{n-1}$.
\end{Theorem}

The large genus asymptotic behaviors of the number of ribbon graphs with a fixed number of faces are discovered in~\cite{PC}.

\begin{Theorem}\label{thm2RG}
	For fixed $n\ge1$ and fixed integers $i_1,\dots,i_{n-1}\ge1$, there exists a rational function $R^{\rg}(g;i_1,\dots,i_{n-1})$ of $g$ that may depend on $n,i_1,\dots,i_{n-1}$, such that
	\[
		\mathcal{G}^{\rg}_{i_1,\dots,i_{n-1},4g+2n-2-|i|}(g)=R^{\rg}(g;i_1,\dots,i_{n-1}).
	\]
	\end{Theorem}

	Theorems~\ref{thmRG} and~\ref{thm2RG} yield the following corollary.
\begin{Corollary}	
	For fixed $n\ge1$ and fixed integers $i_1,\dots,i_{n-1}\ge1$, we have the following convergent expansion:
	\begin{equation}
		\mathcal{G}^{\rg}_{i_1,\dots,i_{n-1},4g+2n-2-|i|}(g)\sim\sum_{k=0}^{\infty}\frac{\mathcal{G}^{\rg}_k}{g^k},\qquad g\to\infty,\label{rgexp}
	\end{equation}
	where $\mathcal{G}^{\rg}_0=1$ and $\mathcal{G}^{\rg}_k=\mathcal{G}^{\rg}_k(n,i_1,\dots,i_{n-1})$, $k\ge1$, are constants.
\end{Corollary}

The rest of the paper is organized as follows. In Section~\ref{secog}, we prove Theorems~\ref{thmOG} and~\ref{thm2OG}. In Section~\ref{secrg}, we prove Theorems~\ref{thmRG} and~\ref{thm2RG}.

\section{Asymptotic formula on enumeration of ordinary graphs} \label{secog}
In this section, we prove Theorems~\ref{thmOG} and~\ref{thm2OG}.
As in~\cite{GYPsi}, we consider the rational function
\[
	P(\sigma;\lambda_1,\dots,\lambda_n)=\frac{1}{\prod_{q=1}^{n}\bigl(\lambda_{\sigma(q)}-\lambda_{\sigma(q+1)}\bigr)}.
\]
The formal Laurent series of the whole right-hand side of~\eqref{main2} is independent of the ordering. Thus, following~\cite{GYPsi}, we choose the region of expansion to be $|\lambda_1|>\dots>|\lambda_n|$. We have the following lemma~\cite{GYPsi}.
\begin{Lemma}[Guo--Yang \cite{GYPsi}] \label{LaurentP}
	For each $\sigma\in S_n$, the Laurent expansion of the rational function $P(\sigma;\lambda_1,\dots,\lambda_n)$ around~$\infty$'s within the region $|\lambda_1|>\dots>|\lambda_n|$ is given by
	\[
		P(\sigma;\lambda_1,\dots,\lambda_n) = (-1)^{m(\sigma)} \sum_{j_1,\dots,j_n\geq0} \prod_{q=1}^n
		\lambda_{\sigma(q)}^{J_{\sigma,q}(j_q)-J_{\sigma,q-1}(j_{q-1})-1}.
	\]
	Here,
	\[
		m(\sigma):= {\rm card} \{q \in \{1,\dots,n\} \mid \sigma(q+1)<\sigma(q) \},
	\]
	and
	\begin{equation}
		J_{\sigma,q}(j) :=
		 \begin{cases}
			-j-1, &\sigma(q)<\sigma(q+1), \\
			j, &\sigma(q)>\sigma(q+1),
		\end{cases}\label{defjsi}
	\end{equation}
	where the indices are considered cyclically modulo $n$.
\end{Lemma}

For $n\ge2$, \smash{$\underline{d}=(d_1,\dots,d_n)\in\bigl(\ZZ^{\ge0}\bigr)^n$}, $\sigma\in S_n$, and $q=1,\dots,n$, we define \smash{$K_{\underline{d}, \sigma, q}\colon \bigl(\mathbb{Z}^{\ge 0}\bigr)^n\rightarrow \mathbb{Z}$} by
\begin{equation}
	K_{\underline{d}, \sigma, q} (\underline{j}):=d_{\sigma(q)}+J_{\sigma,q}(j_q)-J_{\sigma,q-1}(j_{q-1}), \label{defK}
\end{equation}
where \smash{$\underline{j}=(j_1,\dots,j_n) \in \bigl(\mathbb{Z}^{\ge 0}\bigr)^n$}, and $J_{\sigma,q}$ are given in~\eqref{defjsi}.

For $n\ge 2$, $1\le r\le n-1$ being integers, we call a permutation $\sigma\in S_n$ with $\sigma(n)=n$ an~{\it $(r,n-r)$-permutation}, if
\begin{align*}
	& n=\sigma(n)>\sigma(1)>\dots >\sigma(r)=1,\\
	& 1=\sigma(r)<\sigma(r+1)<\cdots <\sigma(n)=n
\end{align*}
(namely, {\it firstly decreasing then increasing}, often called {\it unimodal} permutations). The set of all permutations of this type is denoted by \smash{$S_{n,n}^{(r,n-r)}$}. We also denote
\smash{$S_{n,n}^{\{2\}}=\bigcup_{r=1}^{n-1}S_{n,n}^{(r,n-r)}$}. The following combinatorial facts will be used later.
\begin{align*}
	& \operatorname{card} S_{n,n}^{(r,n-r)}=\binom{n-2}{r-1},\\
	& \operatorname{card} S_{n,n}^{\{2\}}=2^{n-2}.
\end{align*}

For $k\ge0$, define a sequence of matrices $L_k$ by
\[
	L_k:=
	\begin{cases}
		\begin{pmatrix}
			1&0\\0&0
		\end{pmatrix}, & k=0,\vspace{1mm}\\
		\begin{pmatrix}
			(k-1)!!&0\\0&-(k-1)!!
		\end{pmatrix}, & k\text{ even and }k\ge2,\vspace{1mm}\\
		\begin{pmatrix}
			0&-k!!\\(k-2)!!&0
		\end{pmatrix}, & k\text{ odd and }k\ge1.
	\end{cases}
\]
Then, by definition,
\[
	R_1(\lambda)=\sum_{k\ge0}L_k\lambda^{-k}.
\]
For $k_1,\dots,k_n\ge0$, we also denote by
\begin{equation}
	l_{k_1,\dots,k_n}=\operatorname{tr}(L_{k_1}\cdots L_{k_n}). \label{defl}
\end{equation}
We make the convention that $l_{k_1,\dots,k_n}=0$ if some of $k_1,\dots,k_n$ are less than or equal to $-1$. Note that $l_{k_1,\dots,k_n}$ is invariant under cyclic permutations of its indices.

With the above notations, we obtain an explicit formula on the normalized number of ordinary graphs by using Lemma~\ref{LaurentP} and expanding the right-hand side of~\eqref{main2}, which is given by the following lemma.
\begin{Lemma}\label{lem2}
	For $n\ge2$, $g\ge0$ and \smash{$\underline{i}=(i_1,\dots,i_n)\in\bigl(\mathbb{Z}^{\ge1}\bigr)^n$} satisfying $i_1+\cdots+i_n=2g+2n-2$, the following formula holds true:
	\begin{equation}
		\mathcal{G}^{\og}_{\underline{i}}(g)=\frac{1}{(2g+2n-3)!!}\sum_{\substack{\sigma\in S_n\\\sigma(n)=n}}(-1)^{m(\sigma)+1}\sum_{\underline{j}\in(\mathbb{Z}^{\ge0})^n}l_{K_{\underline{i},\sigma,1}(\underline{j}),\dots,K_{\underline{i},\sigma,n}(\underline{j})}. \label{ogmidmain}
	\end{equation}
\end{Lemma}
Due to our convention, for each $\sigma\in S_n$, the second sum on the right-hand side of~\eqref{ogmidmain} is a finite sum. Moreover, when $i_1,\dots,i_{n-1}$ are fixed and $g$ is sufficiently large, the number of terms in the sum is independent of $g$. Also note that a cyclic permutation of $\sigma$ does not change the value of the summand of the right-hand side of~\eqref{main2}. Thus we may restrict the permutations $\sigma$ to $\sigma(n)=n$ and remove the factor $1/n$ appearing in~\eqref{main2}. Now we are ready to prove Theorems~\ref{thmOG} and~\ref{thm2OG}.

\begin{proof}[Proof of Theorem~\ref{thmOG}]
	For the case $n=1$, we know from~\eqref{main1} that $\mathcal{G}^{\og}_{2g}(g)=1$ for every $g\ge1$. So we consider the case $n\ge2$. For fixed integers $i_1,\dots,i_{n-1}\ge1$, let $i_n=2g+2n-2-|i|$. From the definition~\eqref{defK} of $K_{\underline{i},\sigma,q}$, for each $\sigma\in S_n$ satisfying $\sigma(n)=n$, we have
	\begin{align}
		& K_{\underline{i},\sigma,n}\ge2g+2n-2-|i|, \label{ogK1}\\
		& K_{\underline{i},\sigma,1}+\cdots+K_{\underline{i},\sigma,n}=2g+2n-2. \label{ogK2}
	\end{align}
	For simplicity, we write $\underline{K}=(K_{\underline{i},\sigma,1}(\underline{j}),\dots,K_{\underline{i},\sigma,n}(\underline{j}))$. Then from the definition of $l_{i_1,\dots,i_n}$, \eqref{ogK1}~and~\eqref{ogK2}, we have for $n\ge1$
	\begin{equation}
		\lim\limits_{g\to\infty}\frac{1}{(2g+2n-3)!!}l_{\underline{K}}=
		\begin{cases}
			1,&\underline{K}=(0,0,\dots,0,0,2g+2n-2),\\
			-1,&\underline{K}=(1,0,\dots,0,0,2g+2n-3),\\
			0,&\text{otherwise}.
		\end{cases} \label{lim}
	\end{equation}
	Now we need to respectively count the number of solutions to the two equations
	\begin{align}
		&\underline{K}=(0,0,\dots,0,0,2g+2n-2),\label{case1}\\
		&\underline{K}=(1,0,\dots,0,0,2g+2n-3)\label{case2}
	\end{align}
	for \smash{$\underline{j}\in\bigl(\mathbb{Z}^{\ge0}\bigr)^n$}.
	
	For~\eqref{case1}, a solution $\underline{j}$ satisfies that
	\[
		J_{\sigma,q}(j_q)-J_{\sigma,q-1}(j_{q-1})=-i_{\sigma(q)},\qquad 1\le q\le n-1.
	\]
	Thus the sequence $\{J_{\sigma,n}(j_n),J_{\sigma,1}(j_1),\dots,J_{\sigma,n-1}(j_{n-1})\}$ is decreasing, which means that \smash{$\sigma\!\in\! S_{n,n}^{\{2\}}$}. For \smash{$\sigma\in S_{n,n}^{(r,n-r)}$}, \smash{$J_{\sigma,r-1}(j_{r-1})\ge0$} and \smash{$J_{\sigma,r}(j_r)\le-1$}. Thus
	\[
		\sum_{q=1}^{r-1}i_{\sigma(q)}\le J_{\sigma,n}(j_n)\le\sum_{q=1}^r i_{\sigma(q)}-1.
	\]
	Therefore, we see that the number of solutions for $J_{\sigma,n}(j_n)$, which is the same as the number of solutions for $\underline j$, is $i_1$.

	For~\eqref{case2}, a solution $\underline j$ satisfies that
	\[
		J_{\sigma,q}(j_q)-J_{\sigma,q-1}(j_{q-1})=\delta_{q,1}-i_{\sigma(q)},\qquad 1\le q\le n-1.
	\]
	Similarly, if the unimodal permutation \smash{$\sigma\in S_{n,n}^{(r,n-r)}$}, $r\neq1$, we see that the number of solutions for $\underline j$ is $i_1$. If \smash{$\sigma\in S_{n,n}^{(1,n-1)}$}, we have
	\[
		0\le J_{\sigma,n}(j_n)\le i_1-2,
	\]
	which means that the number of solutions for $\underline j$ is $i_1-1$.

	Therefore, from~\eqref{lim}, Lemma~\ref{lem2} and the above statement, we have
	\begin{align*}
		&\lim\limits_{g\to\infty}\mathcal{G}^{\og}_{i_1,\dots,i_{n-1},2g+2n-2-|i|}(g)\\
		&\qquad{}=\lim\limits_{g\to\infty}\frac{1}{(2g+2n-3)!!}\sum_{\substack{\sigma\in S_n\\\sigma(n)=n}}(-1)^{m(\sigma)+1}\sum_{\underline{j}\in(\mathbb{Z}^{\ge0})^n}l_{K_{\underline{i},\sigma,1}(\underline{j}),\dots,K_{\underline{i},\sigma,n}(\underline{j})}\\
		&\qquad{}=\lim\limits_{g\to\infty}\frac{1}{(2g+2n-3)!!}\sum_{\sigma\in S_{n,n}^{\{2\}}}(-1)^{m(\sigma)+1}\sum_{\underline{j}\in(\mathbb{Z}^{\ge0})^n}l_{K_{\underline{i},\sigma,1}(\underline{j}),\dots,K_{\underline{i},\sigma,n}(\underline{j})}\\
		&\qquad{}=\lim\limits_{g\to\infty}\frac{1}{(2g+2n-3)!!}\sum_{r=1}^{n-1}\sum_{\sigma\in S_{n,n}^{(r,n-r)}}(-1)^{r+1}\sum_{\underline{j}\in(\mathbb{Z}^{\ge0})^n}l_{K_{\underline{i},\sigma,1}(\underline{j}),\dots,K_{\underline{i},\sigma,n}(\underline{j})}\\
		&\qquad{}=\lim\limits_{g\to\infty}\sum_{r=1}^{n-1}(-1)^{r+1}\binom{n-2}{r-1}(i_1-(i_1-\delta_{1,r}))\\
		&\qquad{}=1.
	\end{align*}
	Here, we recall that since $i_1,\dots,i_{n-1}$ are fixed, the number of terms in the right-hand side of~\eqref{ogmidmain} is a constant independent of $g$ for sufficiently large $g$. Then the theorem is proved.
\end{proof}

Now let us proceed to prove Theorem~\ref{thm2OG}.

\begin{proof}[Proof of Theorem~\ref{thm2OG}]
	For $n=1$, the statement easily follows from~\eqref{main1}. Now consider the case $n\ge2$. For fixed integers $i_1,\dots,i_{n-1}$ and $i_n=2g+2n-2-|i|$, using~\eqref{ogK1},~\eqref{ogK2} and~\eqref{defl}, one can see that for each $\sigma\in S_n$, \smash{$\frac{1}{(2g+2n-3)!!}l_{\underline{K}}$} is a rational function of $g$. Since the number of the terms in the right-hand side of~\eqref{ogmidmain} is independent of $g$ for sufficiently large~$g$, we conclude from~\eqref{ogmidmain} the existence of a rational function $R^{\og}(g;i_1,\dots,i_{n-1})$ such that ${\mathcal{G}^{\og}_{i_1,\dots,i_n}(g)=R^{\og}(g;i_1,\dots,i_{n-1})}$.
\end{proof}

\section[Asymptotic formula on enumeration of ribbon graphs with 1 face]{Asymptotic formula on enumeration\\ of ribbon graphs with 1 face} \label{secrg}
In this section, we prove Theorems~\ref{thmRG} and~\ref{thm2RG}.
Define a sequence of matrix-valued polynomials $B_k$ in $N$ for each $k\ge0$ by
	\[
		B_k=\begin{cases}
			\begin{pmatrix}
				1&0\\0&0
			\end{pmatrix}, & k=0,\vspace{1mm}\\
			\begin{pmatrix}
				N(k-1)!!\cdot\frac{1+(-1)^{(k-2)/2}}{k}&0\\0&-N(k-1)!!\cdot\frac{1+(-1)^{(k-2)/2}}{k}
			\end{pmatrix}, & k \ \text{even and} \ k\ge2,\vspace{1mm}\\
			\begin{pmatrix}
				0&-N(k-2)!!\\(k-2)!!\cdot(-1)^{(k-1)/2}&0
			\end{pmatrix}, &k \ \text{odd and} \ k\ge1.
		\end{cases}
	\]
For $k_1,\dots,k_n\ge0$, we denote
\begin{equation}
	b_{k_1,\dots,k_n}=\bigl[N^1\bigr] \operatorname{tr}(B_{k_1}\cdots B_{k_n}).\label{defb}
\end{equation}
We make the convention that $b_{k_1,\dots,k_n}=0$ if some of $k_1,\dots,k_n$ are less than or equal to $-1$. Note that $b_{k_1,\dots,k_n}$ is invariant under cyclic permutations of its indices. Also note that since the $N^0$ parts of $R_N$ only appear in the left column, the $N^1$ part in the lower-left entry of $R_N$ for $k$ odd does not contribute to the first order coefficient of the trace and therefore can be omitted. With these definitions, we obtain an explicit formula for the normalized number of ribbon graphs by using Lemma~\ref{LaurentP} and expanding the right-hand side of~\eqref{main2}, which is given by the following lemma.
	\begin{Lemma}\label{lemRG}
		For $n\ge2$, $g\ge0$ and $\underline{i}=(i_1,\dots,i_n)\in\bigl(\mathbb{Z}^{\ge1}\bigr)^n$ satisfying $i_1+\cdots+i_n=4g+2n-2$, the following formula holds true:
		\begin{equation}
			\mathcal{G}^{\rg}_{\underline{i}}(g)=\frac{1}{2\cdot(4g+2n-5)!!}\sum_{\substack{\sigma\in S_n\\\sigma(n)=n}}(-1)^{m(\sigma)+1}\sum_{\underline{j}\in(\mathbb{Z}^{\ge0})^n}b_{K_{\underline{i},\sigma,1}(\underline{j}),\dots,K_{\underline{i},\sigma,n}(\underline{j})}. \label{midmain}
		\end{equation}
\end{Lemma}
By our convention, for each $\sigma\in S_n$ the second sum on the right-hand side of~\eqref{midmain} is a finite sum. Moreover, when $i_1,\dots,i_{n-1}$ are fixed and $g$ is sufficiently large, the number of terms in the sum is independent of $g$. Now we are ready to prove Theorems~\ref{thmRG} and~\ref{thm2RG}.

\begin{proof}[Proof of Theorem~\ref{thmRG}]
		For the case $n=1$, we know from~\eqref{main1} that $\mathcal{G}^{\rg}_{4g}(g)=1$ for every ${g\ge1}$. Now we consider the case $n\ge2$. For fixed integers $i_1,\dots,i_{n-1}\ge1$, let $i_n=4g+2n-2-|i|$. From the definition~\eqref{defK} of $K_{\underline{i},\sigma,q}$, for each $\sigma\in S_n$ satisfying $\sigma(n)=n$, we have
		\begin{align}
			& K_{\underline{i},\sigma,n}\ge4g+2n-2-|i|, \label{K1}\\
			& K_{\underline{i},\sigma,1}+\cdots+K_{\underline{i},\sigma,n}=4g+2n-2. \label{K2}
		\end{align}
		For simplicity we write $\underline{K}=(K_{\underline{i},\sigma,1}(\underline{j}),\dots,K_{\underline{i},\sigma,n}(\underline{j}))$. Then from the definition of $b_{k_1,\dots,k_n}$,~\eqref{K1} and~\eqref{K2}, we have for $n\ge2$
		\begin{equation}
			\lim\limits_{g\to\infty}\frac{1}{2\cdot(4g+2n-5)!!}b_{\underline{K}}=
			\begin{cases}
				\frac{1}{2}(1+(-1)^n),&\underline{K}=(0,0,\dots,0,0,4g+2n-2),\\
				-\frac{(-1)^n}{2},&\underline{K}=(0,0,\dots,0,1,4g+2n-3),\\
				-\frac{1}{2},&\underline{K}=(1,0,\dots,0,0,4g+2n-3),\\
				0,&\text{otherwise}.
			\end{cases} \label{limRG}
		\end{equation}
		Now we need to respectively count the number of solutions to the three equations
		\begin{align}
			&\underline{K}=(0,0,\dots,0,0,4g+2n-2),\label{case1RG}\\
			&\underline{K}=(0,0,\dots,0,1,4g+2n-3),\label{case2RG}\\
			&\underline{K}=(1,0,\dots,0,0,4g+2n-3)\label{case3RG}
		\end{align}
		for $\underline{j}\in(\mathbb{Z}^{\ge0})^n$. Using the arguments similar to the proof of Theorem~\ref{thmOG}, it turns out that the number of solutions to~\eqref{case1RG} for $\underline{j}$ is
		\[
			\begin{cases}
				i_1,&\sigma\in S_{n,n}^{\{2\}},\\
				0,&\sigma\not\in S_{n,n}^{\{2\}},
			\end{cases}
		\]
		the number of solutions to~\eqref{case2RG} for $\underline{j}$ is
		\[
			\begin{cases}
				i_1-1,&\sigma\in S_{n,n}^{(n-1,1)},\\
				i_1,&\sigma\in S_{n,n}^{(r,n-r)}\text{ with }r\not=n-1,\\
				0,&\sigma\not\in S_{n,n}^{\{2\}},
			\end{cases}
		\]
		and the number of solutions to~\eqref{case3RG} for $\underline{j}$ is
		\[
			\begin{cases}
				i_1-1,&\sigma\in S_{n,n}^{(1,n-1)},\\
				i_1,&\sigma\in S_{n,n}^{(r,n-r)}\text{ with }r\not=1,\\
				0,&\sigma\not\in S_{n,n}^{\{2\}}.
			\end{cases}
		\]
		Therefore, from~\eqref{limRG}, Lemma~\ref{lemRG} and the above statement, we have
		\begin{align*}
			&\lim\limits_{g\to\infty}\mathcal{G}^\rg_{i_1,\dots,i_{n-1},4g+2n-2-|i|}(g)\\
			&\qquad{}=\lim\limits_{g\to\infty}\frac{1}{2\cdot(4g+2n-5)!!}\sum_{\substack{\sigma\in S_n\\\sigma(n)=n}}(-1)^{m(\sigma)+1}\sum_{\underline{j}\in(\mathbb{Z}^{\ge0})^n}b_{K_{\underline{i},\sigma,1}(\underline{j}),\dots,K_{\underline{i},\sigma,n}(\underline{j})}\\
			&\qquad{}=\lim\limits_{g\to\infty}\frac{1}{2\cdot(4g+2n-5)!!}\sum_{\sigma\in S_{n,n}^{\{2\}}}(-1)^{m(\sigma)+1}\sum_{\underline{j}\in(\mathbb{Z}^{\ge0})^n}b_{K_{\underline{i},\sigma,1}(\underline{j}),\dots,K_{\underline{i},\sigma,n}(\underline{j})}\\
			&\qquad{}=\lim\limits_{g\to\infty}\frac{1}{2\cdot(4g+2n-5)!!}\sum_{r=1}^{n-1}\sum_{\sigma\in S_{n,n}^{(r,n-r)}}(-1)^{r+1}\sum_{\underline{j}\in(\mathbb{Z}^{\ge0})^n}b_{K_{\underline{i},\sigma,1}(\underline{j}),\dots,K_{\underline{i},\sigma,n}(\underline{j})}\\
			&\qquad{}=\biggl(\biggl(\frac{1}{2}(1+(-1)^n)-\frac{(-1)^n}{2}\biggr)i_1-\frac{1}{2}(i_1-1)\biggr)\\
			&\qquad\quad{}+\sum_{r=2}^{n-2}(-1)^{r+1}\binom{n-2}{r-1}\biggl(\frac{1}{2}(1+(-1)^n)-\frac{(-1)^n}{2}-\frac{1}{2}\biggr)i_1\\
			&\qquad\quad{}+(-1)^{n}\biggl(\biggl(\frac{1}{2}(1+(-1)^n)-\frac{1}{2}\biggr)i_1-\frac{(-1)^n}{2}(i_1-1)\biggr)\\
			&\qquad{}=1,
		\end{align*}
		which proves Theorem~\ref{thmRG}.
\end{proof}
	
We note that an upper bound for correlators in topological recursion was given in~\cite{BEG} in a~general context. In our situation, the upper bound given in~\cite{BEG} involves a coefficient with an~unspecified dependence on the genus $g$; on the other hand, our estimate here gives the precise leading asymptotics.
	
	\begin{proof}[Proof of Theorem~\ref{thm2RG}]
		For $n=1$, the statement easily follows from~\eqref{main1}. Now consider the case $n\ge2$. For fixed integers $i_1,\dots,i_{n-1}$ and $i_n=4g+2n-2-|i|$, using~\eqref{K1},~\eqref{K2} and~\eqref{defb}, one can see that for each $\sigma\in S_n$, \smash{$\frac{1}{2\cdot(4g+2n-5)!!}b_{\underline{K}}$} is a rational function of $g$. Since the number of the terms in the right-hand side of~\eqref{midmain} is independent of $g$ for sufficiently large~$g$, we conclude from~\eqref{midmain} the existence of a rational function $R^{\rg}(g;i_1,\dots,i_{n-1})$ such that $\mathcal{G}^{\rg}_{i_1,\dots,i_n}(g)=R^{\rg}(g;i_1,\dots,i_{n-1})$.
\end{proof}

We note that the matrix-resolvent formulae~\eqref{main1} and~\eqref{main2} not only give the rationality as shown in Theorems~\ref{thm2OG} and~\ref{thm2RG}, but also lead to efficient computations for the associated rational functions (cf.~\cite{BDY1,GNYZ,GYPsi,GYZ}). Moreover, similar to~\cite{GYPsi}, the expressions of $\mathcal{G}_k^\og$ and~$\mathcal{G}_k^\rg$ in~\eqref{ogexp} and~\eqref{rgexp} for~$k$ small can then be computed. Recall that for intersection numbers a~polynomiality conjecture of such coefficients was proposed in~\cite{GYPsi} (cf.~\cite{LX}), which was proved in~\cite{Eynard,GYZ} (cf.~\cite{GNYZ}); similar polynomiality for $\mathcal{G}_k^\og$ and $\mathcal{G}_k^\rg$ is found in~\cite{PC}.

\subsection*{Acknowledgements}

The author would like to thank Di Yang for his advice, suggestions of the questions, and helpful discussions. The author is also grateful to Jindong Guo for helpful discussions. He thanks the anonymous referees for very helpful suggestions. Part of the work was done while the author was an undergraduate student of USTC; he thanks USTC for excellent working conditions.

\pdfbookmark[1]{References}{ref}
\LastPageEnding

\end{document}